\documentclass[12pt]{article}
\usepackage{graphicx}
\usepackage[cp1251]{inputenc}
\begin{document}
\pagestyle{empty}

\pagestyle{headings} {\title{Neutrino mixings as a source of
lepton flavor violations}}
\author{O.M.Boyarkin\thanks{E-mail:oboyarkin@tut.by},
G.G.Boyarkina, D.S.Vasileuskaya\\
\small{\it{Belorussian State University,}}\\
\small{\it{Dolgobrodskaya Street 23, Minsk, 220070, Belarus}}}
\date{}
\maketitle

\begin{abstract}
Within the left-right symmetric model (LRM) the $Z$ boson decay
into the channel $Z\to\tau\mu$ are investigated. The branching
ratios of this decay is found in the third order of the
perturbation theory. The obtained expression does not equal to
zero only at the existence of the neutrino mixings. It means that
from the point of view of the LRM nonconservation both of neutral
and of charged lepton flavors has the same nature. As a result,
elucidation of the decays $Z\to\l_i\overline{l}_k$ ($i\neq k$)
could provide data concerned the neutrino sector structure of the
LRM. The neutrino sector parameters which could be measured in
that case are as follows: (i) difference of the heavy neutrino
masses; (ii) heavy-heavy neutrino mixing; (iii) heavy-light
neutrino mixing.
\end{abstract}
\hspace{5mm}

\hspace*{-6mm}{\it{Keywords}}: $Z$ boson decays, charged lepton
flavor violation, left-right symmetric model, heavy and light
neutrinos,
mixing in the neutrino sector, Large Hadron Collider.\\[5mm]

\section{Introduction}
The standard model (SM) of particle physics has been very
successfully predicting or explaining most experimental results
and phenomena. However it still has a few outstanding problems
with empirical observations. One of them is connected with
neutrinos. In the SM the lepton flavors $L_{e,\mu,\tau}$ are the
conserved quantities. However, neutrino oscillation experiments
demonstrated that the neutrinos have the masses and the neutral
lepton flavors (NLF's)is not conserved. It should be stressed that
this nonconservation is caused by the mixing in the neutrino
sector. Of course, the minimally extended SM (SM with the massive
neutrinos) may be invoked for description of neutrino oscillation
experiments but processes involving violation of charged lepton
flavors (CLF's) are extremely suppressed in it because of the
small neutrino masses. Owing to a positive signal in any of the
experimental looking for CLF violation (CLFV) processes would
automatically imply the existence of physics beyond the SM.
Although no such processes have been detected to date, this is a
very active field that is being explored by many experiments which
have adjusted upper limits to this kind of CLFV processes.

The CLFV processes can be classified into high energy ones that
are detected at colliders, such as the CLFV decays of the $Z$ and
Higgs bosons, and low energy ones such as $\mu - e$ conversion in
nuclei, rare radiative and pure leptonic decays of the $\mu$ and
$\tau$ leptons. By now the strongest limits on the CLFV processes
have been set in the $\mu-e$ conversions. For example, the
branching ratios of the radiative $\mu\to e\gamma$ decay and
$\mu-e$ transition in heavy nuclei have been bounded to be below
$4.2\times10^{-13}$ and $7.0\times10^{-13}$ by the MEG
\cite{AMB16} and SINDRUM II \cite{WHB} collaborations,
respectively. Next generation of experiments are expected to
enhance in several orders of magnitude the sensitivities for
$\mu-e$ transitions, reaching the impressive range of $10^{-18}$
for $\mu-e$ transition in nuclei by the PRISM experiment in J-PARC
\cite{AA13}.

LEP, as $Z$ factory, looked for the CLFV decays $Z\to
l_i\overline{l}_k$ ($i\neq k$) with no luck. In such a manner it
established upper limits to these processes, which are relatively
weak compared with the low-energy processes
$$\mbox{BR}(Z\to
e\mu)<1.7\times10^{-6}\qquad \cite{LEP1},\eqno(1)$$
$$\mbox{BR}(Z\to e\tau)<9.8\times10^{-6},\qquad
\cite{LEP1},
\eqno(2)$$
$$\mbox{BR}(Z\to \mu\tau)<1.2\times10^{-5}\qquad \cite{LEP2}.
\eqno(3)$$

The currently running LHC could also throw light on CLFV
processes. The LHC has been searching the $Z$ boson decays into
two leptons of different flavor as well \cite{GA14,ATLAS3}. ATLAS
is already at the level of LEP results for the LFV $Z$ decay
rates, and even better for $Z\to\mu e$ channel
$$\mbox{BR}(Z\to
e\mu)<7.50\times10^{-7}\qquad \cite{ATLAS3},\eqno(4)$$
$$\mbox{BR}(Z\to \mu\tau)<1.3\times10^{-5}\qquad \cite{GA14},
\eqno(5)$$
$$\mbox{BR}(Z\to e\tau)<5.8\times10^{-5}\qquad \cite{GA14}.
\eqno(6)$$

The CLFV is also investigated in the Higgs boson decays $H\to
l_k\overline{l}_m$, which are searched by the CMS
\cite{VK15,CMS17} and ATLAS \cite{ATLAS3} collaborations. There
are a lot of models predicting the CLFV in the decays both of the
Higgs \cite{KC16,SB16,Eah16,AA16} and $Z$ bosons
\cite{AAV15,MAP4,SDS12}. It is clear that amongst them the models
having common mechanism both for NLF violation and for CLFV are
most attractive. The left-right model (LRM) \cite{ICP74} belongs
among such models. The neutrino sector of the LRM, apart from
light left-handed neutrinos $\nu_{lL}$, also includes heavy
right-handed neutrinos $N_{lR}$ which are partners on the see-saw
mechanism for $\nu_{lL}$. As this takes place, mixings in the
neutrino sector become a principal source of the CLFV.

Within the LRM the CLFV has already been examined in Refs.
\cite{Bom97,Bom04,Bom18}. The goal of our work is to consider the
CLFV decays of the $Z$ boson and establish what parameters of the
LRM neutrino sector therewith could be determined. In the next
chapter we give a short summary of the LRM (the detail description
of the model could be found in the book \cite{OMB11}). In sections
3 we calculate the branching ratio of the decay $Z\to\tau^-\mu^+$
in the third order of the perturbation theory. Our results are
discussed in section 4.

\section{The left-right-symmetry and neutrino mixing}
The LRM is built upon the gauge group $SU(2)_L\times SU(2)_R\times
U(1)_{B-L}$. It has three gauge coupling constants $g_L, g_R$ and
$g^{\prime}$ for the $SU(2)_L$, $SU(2)_R$ and $U(1)_{B-L}$ groups,
respectively. In the LRM quarks and leptons appears in the left-
and right-handed doublets
$$\left.\begin{array}{ll}
\displaystyle{Q_L^{\alpha}({1\over2}, 0,
{1\over3})=\left(\matrix{u_L^{\alpha}\cr
d_L^{\alpha}}\right)},\hspace{12mm}
\displaystyle{Q_R^{\alpha}(0,{1\over2},
{1\over3})=\left(\matrix{u_R^{\alpha}\cr
d_R^{\alpha}}\right)},\\[4mm]
\displaystyle{\Psi_L^a({1\over2}, 0, -1)=\left(\matrix{\nu_{a
L}\cr l_{\alpha L}}\right)},\qquad
\displaystyle{\Psi_R^a(0,{1\over2}, -1)=\left(\matrix{N_{a R}\cr
l_{a R}}\right)},\end{array}\right\}\eqno(7)$$ where in brackets
the values of $S^W_L, S^W_R$ and $B-L$ are given, $S^W_L$
($S^W_R$) is the weak left (right) isospin, $\alpha=\mbox{red},
\mbox{blue},\mbox{green}$, and $a=e,\mu,\tau$. The scalar sector
of the LRM, as a rule, contains the bi-doublet $\Phi(1/2,1/2,0)$
and two triplets $\Delta_L(1,0,2)$, $\Delta_R(0,1,2)$ and, as a
result, neutrinos are Majorana particles.

After spontaneous symmetry breaking which is realized by the
following choice of the vacuum expectation values (VEV's)
$$\begin{array}{ll}<\Delta^0_{L,R}>=v_{L,R}/\sqrt{2},
\qquad <\Phi^0_1>=k_1,\qquad <\Phi^0_2>=k_2,\\[2mm]
\hspace*{30mm}v_L<<\mbox{max}(k_1,k_2)<<v_R,\end{array}\eqno(8)$$
the gauge boson sector include two neutral ($Z_{1,2}$) and two
charged ($W_{1,2}$) gauge bosons, where $Z_1$ and $W_1$ bosons are
analogs of the $Z$ and $W$ bosons of the SM, respectively.

The Lagrangian describing interaction of the charged gauge bosons
with the $Z_{1,2}$ bosons is conveniently expressed by the
following form \cite{OMB11}
$${\cal L}_{WWV}=i\rho^{(V)}_{kl}B_{\mu\nu,\lambda\sigma}\Big\{[\partial^{\mu}
W^{\lambda*}_k(x)]W^{\nu}_l(x)V^{\sigma}(x)+W^{\sigma*}_k(x)[\partial^{\mu}
W^{\lambda}_l(x)]V^{\nu}(x)+$$
$$+W^{\nu*}_k(x)W^{\sigma}_l(x)[\partial^{\mu}V^{\lambda}(x)]
\Big\},\eqno(9)$$ where $k,l=1,2$, $V=Z_1, Z_2$,
$$\rho^{(Z_1)}_{ll}=\cos^2\left(\xi+\frac{\pi}{2}\delta_{l2}\right) g_LM_{11}+
\sin^2\left(\xi+\frac{\pi}{2}\delta_{l2}\right)
g_RM_{12},\eqno(10)$$
$$\rho^{(Z_1)}_{kl}=\rho^{(Z_1)}_{lk}=\frac{1}{2}\sin{2\xi}(g_LM_{11}-g_RM_{12}),
\qquad k\neq l,\eqno(11)$$
$$W_1=W_L\cos\xi+W_R\sin\xi,\qquad W_2=-W_L\sin\xi+W_R\cos\xi,\qquad
B^{\mu\nu,\lambda\sigma}=g^{\mu\nu}g^{\lambda\sigma}-g^{\mu\sigma}
g^{\nu\lambda},$$ $c_W=\cos\theta_W, \ s_W=\sin\theta_W,$
$\theta_W$ is the Weinberg angle, and $M_{ik}$ are elements of the
matrix
$$M=\left(\matrix{
\cos\phi & \sin\phi\vspace{2mm} \cr  -\sin\phi &
\cos\phi}\right)\left( \matrix{e(g^{\prime-2}+g_R^{-2})^{1/2} &
-eg_L^{-1}g_R^{-1}(g^{\prime-2}+g_R^{-2})^{-1/2}\vspace{2mm}\cr 0
&
g^{\prime-1}(g^{\prime-2}+g_R^{-2})^{-1/2}\cr}\right),\eqno(12)$$
($\phi$ is the mixing angle in the neutral gauge bosons sector).
In its turn the expressions for $\rho^{(Z_2)}_{ll},
\rho^{(Z_2)}_{kl}$ follow from (10) and (11) under the
substitutions
$$g_LM_{11}\rightarrow g_RM_{22}, \qquad g_RM_{12}\rightarrow
g_LM_{21}, \qquad \xi\rightarrow \xi+{\pi\over2}.\eqno(13)$$

The LRM made the following predictions about the values of the
mixings in the gauge bosons sector (see, for example, the book
\cite{OMB11} and references therein)
$$\tan2\phi\simeq{k_+^2(\cos2\theta_W)^{3/2}\over2v_R^2\cos^4\theta_W}\simeq
2{m_Z^2\over m_{Z^{\prime}}^2}\sqrt{\cos2\theta_W},\eqno(14)$$ and
$$\tan2\xi\simeq{4g_Lg_Rk_1k_2\over
g_R^2(2v_R^2+k_+^2)-g_L^2(2v_L^2+k_+^2)},\eqno(15)$$ where
$k_+=\sqrt{k_1^2+k_2^2}=174$ GeV. Further on we shall assume
$g_L=g_R$ and use the designation $g$ for them.

Using the lower bounds on the masses of additional gauge bosons
$$m_{W_2}\geq3.7\ \mbox{TeV}\ (\cite{ATC17}),\qquad
m_{Z^{\prime}}>4.4 \ \mbox{TeV}\ (\cite{TB18}),\eqno(16)$$ and
definitions of the gauge boson masses one could obtain the limits
on these mixing angles. For example, taking into account
$$m_{W_2}^2={1\over2}\Bigg[M_L^2+M_R^2-\sqrt{(M_L^2-M_R^2)^2+4M_{LR}^4}\Bigg]
,$$
where
$$M_L^2={g^2\over2}(k_+^2+2v_L^2),\qquad
M_R^2={g^2\over2}(k_+^2+2v_R^2),\qquad M_{LR}^2=g^2k_1k_2,$$ we
lead to the inequality $v_R\geq5.7$ TeV to give
$$\sin2\xi\leq5\times10^{-4}.\eqno(17)$$
Acting in an analogous way (see, for example, \cite{OMB11}), we
get
$$\tan2\phi<6\times10^{-4}.\eqno(18)$$

In our calculation we also need the Lagrangian which governs the
interaction between charged gauge bosons and fermions
$${\cal{L}}_l^{CC}={1\over2\sqrt{2}}\sum_l\Big[g\overline{l}(x)\gamma^{\mu}
(1-\gamma_5) \nu_{lL}(x)W_{L\mu}(x)+g\overline{l}(x)\gamma^{\mu}
(1+\gamma_5)N_{lR}(x)W_{R\mu}(x)\Big].\eqno(19)$$

The neutrino states entering into the Lagrangian (19) have been
specified in flavor basis. They do not represent physical states
(mass eigenstates), but they are mixing of these states. For the
sake of simplicity, in what follows, we shall be constrained by
two flavor approximation. Then the connection between flavor and
physical basises are determined by the following way
$$\left(\matrix{\nu_{aL}\cr N_{aR}\cr\nu_{bL}\cr N_{bR}}\right)=
\left(\matrix{c_{\varphi_a}c_{\theta_{\nu}} &
s_{\varphi_a}c_{\theta_N} & c_{\varphi_a}s_{\theta_{\nu}}  &
s_{\varphi_a}s_{\theta_N} \cr -s_{\varphi_a}c_{\theta_{\nu}} &
c_{\varphi_a}c_{\theta_N} & -s_{\varphi_a}s_{\theta_{\nu}}  &
c_{\varphi_a}s_{\theta_N} \cr -c_{\varphi_b}s_{\theta_{\nu}}
&-s_{\varphi_b}s_{\theta_N} & c_{\varphi_b}c_{\theta_{\nu}} &
s_{\varphi_b}c_{\theta_N}\cr s_{\varphi_b}s_{\theta_{\nu}}
&-c_{\varphi_b}s_{\theta_N} & -s_{\varphi_b}c_{\theta_{\nu}} &
c_{\varphi_b}c_{\theta_N}\cr}\right)\left(\matrix{\nu_1\cr
N_1\cr\nu_2\cr N_2}\right),\eqno(20)$$ where $\varphi_a$ and
$\varphi_b$ are the mixing angles inside $a$ and $b$ generations
respectively, $\theta_{\nu} (\theta_N)$ is the mixing angle
between the light (heavy) neutrinos belonging to the $a$- and
$b$-generations, $c_{\varphi_a}=\cos\varphi_a, \
s_{\varphi_a}=\sin\varphi_a$ and so on.

Within the LRM one could obtain the exact formula for the
heavy-light neutrino mixing angle $\varphi_{a,b}$ \cite{Bom04}
$$\sin2\varphi_a=2{\sqrt{f^2_{aa}v_Rv_L-[f_{aa}(v_R+v_L)-
m_{\nu_1}c_{\theta_{\nu}}^2-m_{\nu_2}s_{\theta_{\nu}}^2]
(m_{\nu_1}c_{\theta_{\nu}}^2+m_{\nu_2}s_{\theta_{\nu}}^2)} \over
f_{aa}(v_R+v_L)-2(m_{\nu_1}c_{\theta_{\nu}}^2+
m_{\nu_2}s_{\theta_{\nu}}^2)},\eqno(21)$$
$$\sin2\varphi_b=\sin2\varphi_a\left(f_{aa}\rightarrow f_{bb},
\theta_{\nu}\rightarrow\theta_{\nu}+{\pi\over2}\right),\eqno(22)$$
where $f_{aa}$ and $f_{ab}$ are the triplet Yukawa coupling
constants. We see that the heavy-light mixing angles belonging to
different generations are practically equal in value
$$\sin2\varphi_a\simeq\sin2\varphi_b\simeq2{\sqrt{v_Rv_L}\over
v_R+v_L}\equiv\sin2\varphi.\eqno(23)$$

There are a lot of papers devoted to determination of experimental
bounds on the value of the heavy-light neutrino mixing angle
$\varphi$ (see, for example \cite{PSB13} and references therein).
One way to find such bounds is connected with searches for the
neutrinoless double beta decay and disentangle the heavy neutrino
effect. From the results of Ref. \cite{SD16} considering the case
of $^{76}\mbox{Ge}$, it follows that the upper limit on
$\sin\varphi$ is about $few\times10^{-3}$ for $m_N=100$ GeV.

The other way is to directly look for the presence of the
heavy-light neutrino mixing via their signatures, for example, in
collider experiments. By way of illustration, we point Ref.
\cite{CC13} in which the final states with same-sign dileptons
plus two jets without missing energy ($l^{\pm}l^{\pm}jj$),
resulting from $pp$ collisions were considered. Analysis of the
channel
$$p+p\to N^*_ll^{\pm}\to l^{\pm}+l^{\pm}+2j\eqno(24)$$
led to the upper limit on $\sin\varphi$ equal to
$3.3\times10^{-2}$ for $m_{W_R}=4$ TeV and $m_{N_l}=100$ GeV. So
we see that the heavy-light neutrino mixing angle may not be so
small.

In the next chapter, we will show that information about the value
of this angle can also be obtained under investigation of the
decay processes of the $Z$ boson going with the lepton flavor
violation.

\section{CLFV decays of the $Z$ boson}
Let us investigate the $Z_1$ boson decay into the channel
$$Z_1
\to \mu^++\tau^-.\eqno(25)$$ Due to the mixing into the neutrino
sector this decay could proceed in the third order of the
perturbation theory. The corresponding diagrams are shown in
Fig.1.
\begin{figure}[!h]
\begin{center}
\includegraphics[width =0.8\textwidth]{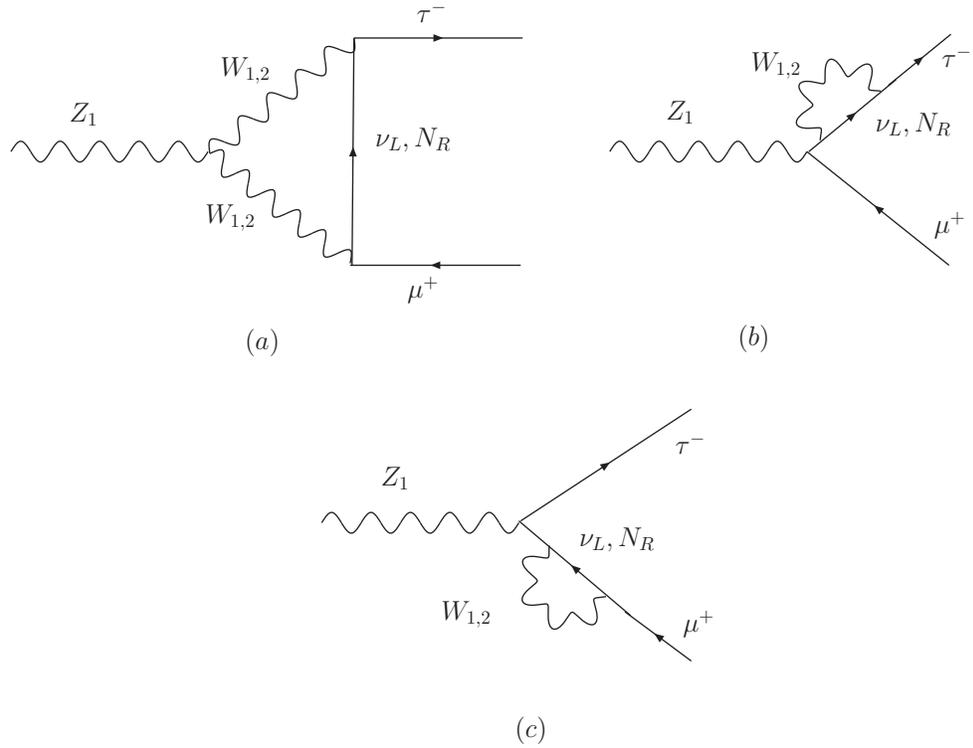}
\end{center}
\caption {The Feynman diagrams contributing to the decay $Z_1\to
\mu^++\tau^-$.}
\end{figure}
For simplicity sake consider the individual contributions of each
diagram to the total width of the decay (25). First we examine the
diagrams shown in Fig.1a. It is clear that the main contribution
to the decay width comes from the diagrams with the
$W_1^+W_1^-\nu_L$ in the virtual state. In this case the internal
neutrino line corresponds to convolution of the operators
$\nu_{\tau L}(x)$ and $\overline{\nu}_{\mu L}(y)$. Then with the
help of Eq. (20) we get
$$\nu_{\tau L}^s(x)\overline{\nu}_{\mu L}^s(y)=
\Big\{-\cos\varphi_{\tau}\sin\theta_{\nu}\nu_1(x)-\sin\varphi_{\tau}\sin
\theta_NN_1(x) +\cos\varphi_{\tau}\cos\theta_{\nu}\nu_2(x)+$$
$$+\sin\varphi_{\tau}\cos\theta_NN_2(x)\Big\}^s
\Big\{\cos\varphi_{\mu}\cos\theta_{\nu}\overline{\nu}_1(y)+
\sin\varphi_{\mu}\cos\theta_N\overline{N}_1(y)+
\cos\varphi_{\mu}\sin\theta_{\nu}\overline{\nu}_2(y)+$$
$$+\sin\varphi_{\mu}\sin\theta_N\overline{N}_2(y)\Big\}^s\simeq
\sin^2\varphi\sin\theta_N\cos\theta_N[N^s_2(x)\overline{N}^s_2(y)-
N^s_1(x)\overline{N}^s_1(y)],\eqno(26)$$ where convolution of the
operators is symbolized by $s$ and we have taken into account
$$\nu^s_1(x)\overline{\nu}^s_1(y)\simeq\nu^s_2(x)\overline{\nu}^s_2(y),\qquad
\varphi_{\mu}=\varphi_{\tau}=\varphi.\eqno(27)$$ The matrix
element corresponding to the diagram under consideration has the
form
$$M^{(a)}={g^3c_W\sin2\theta_{\mu\tau}\sin^2\varphi
\cos\phi\cos^2\xi\over
8}\sqrt{{m_{\tau}m_{\mu}\over2m_{Z_1}E_{\tau}E_{\mu}}}\
\overline{u}(p_1)\gamma^m(1-\gamma_5)\int_{\Omega}\Big\{
\Big[{\hat{k}-{\hat{p}}_2+m_{N_2}\over(k-p_2)^2-m_{N_2}^2}-$$
$$-{\hat{k}-{\hat{p}}_2+m_{N_1}\over(k-p_2)^2-m_{N_1}^2}\Big]
\gamma^n(1-\gamma_5)v(p_2)\Big[g_{\sigma\lambda}\Lambda_{m\nu}(k-p)
\Lambda_{n\beta}(k)k_{\mu}-$$
$$-g_{\nu\lambda}\Lambda_{n\sigma}(k)
\Lambda_{m\beta}(k-p)(k-p)_{\mu}-
g_{\beta\lambda}\Lambda_{m\sigma}(k-p)\Lambda_{n\nu}(k)p_{\mu}\Big]B^{\mu\nu,
\beta\sigma}Z^{\lambda}(p)\Big\}d^4k,\eqno(28)$$ where
$$\Lambda_{\mu\nu}(k)={g_{\mu\nu}-k_{\mu}k_{\nu}/m_{W_1}^2\over
k^2-m_{W_1}^2},$$ $m_{N_j}$ $(j=1,2)$ is the mass of the heavy
neutrino, $p_1$ ($p_2$) is the momentum of $\tau$-lepton
($\mu$-meson), $\theta_{\mu\tau}$ is the mixing angle between the
heavy tau-lepton and muon neutrinos. Thanks to the current upper
limits on the mixing angles in the gauge boson sector we may set
$\cos\phi\cos^2\xi$ equal 1.

The scheme of further calculations is as follows. Using the
procedure of dimensional regularization and considering the motion
equations we rewrite the expression (28) in the form
$$M^{(a)}={i\pi^2g^3c_W\sin2\theta_{\mu\tau}\sin^2\varphi\over
4}\sqrt{{m_{\tau}m_{\mu}\over2m_{Z_1}E_{\tau}E_{\mu}}}\overline{u}
(p_1)\Big[(1+\gamma_5)(A\gamma_{\lambda}+Bp_{1\lambda})+
(1-\gamma_5)(C\gamma_{\lambda}+$$
$$+Dp_{1\lambda})\Big]v(p_2)Z^{\lambda}(p), \eqno(29)$$ where
the quantities $A,B,C$ and $D$ represent the two-dimensional
integrals.

Let us find the part of the partial decay width connected with the
diagram of Fig.1a. Substituting (29) into the formula
$$d\Gamma=(2\pi)^4\delta^{(4)}(p-p_1-p_2)|M^{(a)}|^2{d^3p_1d^3p_2
\over(2\pi)^8},$$ and integrating the obtained expression over
$p_1$, $p_2$, we lead to the result
$$\Gamma(Z_1\to W_1^{-*}W_1^{+*}\nu^*_L\to\tau^-\mu^+)={g^6c_W^2\pi^3
\sin^4\varphi\sin^22\theta_{\mu\tau}
\over384m_{Z_1}^3}\Big\{(m_{Z_1}^2-m_{\mu}^2-m_{\tau}^2)\Big[3f_{A,A}
(m_{N_1},m_{N_2})+$$ $$+3f_{C,C}(m_{N_1},m_{N_2})+\beta_z\Big(
f_{B,B}(m_{N_1},m_{N_2})+f_{D,D}(m_{N_1},m_{N_2})\Big)\Big]+
4f_{A,C}(3m_{\mu}m_{\tau}-2\beta_z)-\beta_ z\Big[4(m_{\mu}+$$
$$+m_{\tau})\Big(f_{C,B}(m_{N_1},m_{N_2})+f_{A,D}(m_{N_1},m_{N_2})\Big)
+4m_{\mu}m_{\tau}f_{B,D}(m_{N_1},m_{N_2})\Big]\Big\}\times$$
$$\times\sqrt{(m_{Z_1}^2-m_{\mu}^2-m_{\tau}^2)^2-
4m_{\mu}^2m_{\tau}^2},\eqno(30)$$ where
$$\beta_z={(m_{Z_1}^2+m_{\tau}^2-m_{\mu}^2)^2\over
m_{Z_1}^2}-m_{\tau}^2,$$
$$f_{A,A}(m_{N_1},m_{N_2})=[A(m_{N_1})-A(m_{N_2})]^2,\qquad f_{A,B}
(m_{N_1},m_{N_2})=[A(m_{N_1})-A(m_{N_2})]\times$$
$$\times[B(m_{N_1})-B(m_{N_2})],$$ and so on. Calculations demonstrate
that the term
$3(m_{Z_1}^2-m_{\mu}^2-m_{\tau}^2)f_{A,A}(m_{N_1},m_{N_2})$
exceeds all remaining terms in the curly brackets on the several
orders of magnitude. Then, taking into account
$$m^2_{Z_1}\gg m^2_{\tau},m^2_{\mu},m_{\tau}m_{\mu}$$
we get
$$\Gamma(Z_1\to W_1^{-*}W_1^{+*}\nu^*_L\to\tau^-\mu^+)\simeq{g^6c_W^2\pi^3\sin^4\varphi\sin^22
\theta_{\mu\tau}m_{Z_1}\over128}
f_{A,A}(m_{N_1},m_{N_2}),\eqno(31)$$ where
$$A(m_j)=\int_0^1dy\int_0^1xdx\Bigg\{\Bigg[8+m_{W_1}^{-2}\Bigg({21\over2}
l_{xy}^j-22p_x^2+
(p_1p_2)(23x-17xy-2)\Bigg)\Bigg]\ln\Bigg|{l_{xy}^j\over
l_{xy}^j-p_x^2}\Bigg|+$$ $$+{1\over
l_{xy}^j-p_x^2}\Bigg[3p_x^2-(p_1p_2)(6x-2xy-4)+m_{W_1}^{-2}\Bigg
(-2p_x^4+p_x^2(p_1p_2)(8x-6xy)-$$
$$-4(p_1p_2)^2(x-xy)(2x-xy)\Bigg)\Bigg]\Bigg\},\eqno(32)$$
$$p_x=p_1(x-xy)+p_2x,\qquad l_{xy}^j=(m_{\mu}^2-m_{N_j}^2-m_{Z_1}^2+
m_{W_1}^2)xy+m_{Z_1}^2x- m_{W_1}^2,$$
$$(p_1p_2)={1\over2}(m_{Z_1}^2-m_{\mu}^2-m_{\tau}^2).$$

Now we embarked on a consideration of contributions of the
diagrams of Fig.1b and Fig.1c. It is clear that these diagrams
transfer to each other under replacement
$$m_{\mu}\longleftrightarrow m_{\tau}.\eqno(33)$$
Therefore, we suffice to examine one of them. After the procedure
of dimensional regularization the matrix element corresponding to
the diagram shown in Fig.1b could be represented in the form
$$M^{(b)}={i\pi^2g^3\sin2\theta_{\mu\tau}\sin^2\varphi\over
16c_W}\sqrt{{m_{\tau}m_{\mu}\over2m_{Z_1}E_{\tau}E_{\mu}}}\
\overline{u}(p_1)\gamma^m(1-\gamma_5)\Big[F(m_{N_2})-F(m_{N_1})\Big]
\gamma^n\times$$ $$\times{\hat{p}_1\over
m_{\tau}^2-m_{\mu}^2}\gamma^{\nu}(\gamma_5-1+4s_W^2)v(p_2)Z_{\nu}(p),
\eqno(34)$$ where
$$F(m_{N_j})=\int_0^1\Big\{g_{mn}\hat{p}_1(1-x)\ln\Big|{l^j_x\over
l^j_x-p_x^2}\Big|-{\hat{p}_1(1-x)\over 2m_{W_1}^2}
\Big[\Big(2p_{xm}p_{xn}+g_{mn}(p_x^2-l_x^j)\Big)\ln\Big|{l^j_x\over
l^j_x-p_x^2}\Big|+g_{mn}p_x^2\Big]+$$
$$+{1\over2m_{W_1}^2}\Big[(p_x^2-l^j_x)
(\gamma_mp_{xn}+\gamma_np_{xm})\ln\Big|{l^j_x\over
l^j_x-p_x^2}\Big|+p_x^2(\gamma_mp_{xn}+
\gamma_np_{xm})\Big]\Big\}dx,\eqno(35)$$
$$p_x=p_1x,\qquad
l_x^j=(m_{\tau}^2+m_{W_1}^2-m_{N_j}^2)x-m_{W_1}^2.$$ Using (34) we
could find $\Gamma(Z_1\to\tau^{-*}W_1^{+*}\nu_L^*\to\tau^-\mu^+)$.
In so doing the following approximate relation takes place
$${\Gamma(Z_1\to\tau^{-*}W_1^{+*}\nu_L^*\to\tau^-\mu^+)\over
\Gamma(Z_1\to
W_1^{-*}W_1^{+*}\nu_L^*\to\tau^-\mu^+)}\simeq10^{-6}\div
10^{-7}.\eqno(36)$$ Therefore, the basic contribution to the decay
(25) is caused by the diagram pictured in Fig.1a.

Let us provide estimation of the branching ratio of the decay
$Z_1\to\tau^-\mu^+$. Before we proceed further, we note that the
function $f_{A,A}(m_{N_1},m_{N_2})$ depends on the difference of
the heavy neutrino masses. For example, when $m_{N_2}$ is varied
from 100 up to 200 Gev and $m_{N_1}=100$ GeV ($m_{N_1}=150$ GeV)
we have
$$f_{A,A}(m_{N_1},m_{N_2})\in [0,0.958],\qquad \Big(f_{A,A}(m_{N_1},m_{N_2})
\in[0,0.125]\Big).\eqno(37)$$ Then setting
$$\theta_{\mu\tau}={\pi\over4},$$
we get
$$\mbox{BR}(Z_1\to\tau\mu)\leq\Bigg\{
\begin{array}{lll}
9.7\times10^{-8},\ \mbox{at}\ \varphi=3.2\times10^{-2},\
m_{N_1}=100\ \mbox{GeV},\ m_{N_2}=150\ \mbox{GeV},\\
[1mm] 1.4\times10^{-8},\ \mbox{at}\ \varphi=5\times10^{-3},\
\ m_{N_1}=100\ \mbox{GeV}, \ m_{N_2}=200\ \mbox{GeV},\\[1mm]
7.6\times10^{-11},\ \mbox{at}\ \varphi=10^{-3},\ \ m_{N_1}=150\
\mbox{GeV}, \ \ m_{N_2}=200\ \mbox{GeV},\end{array}\eqno(38)$$
where
$$\mbox{BR}(Z_1\to\tau\mu)=\mbox{BR}(Z_1\to\tau^-\mu^+)+
\mbox{BR}(Z_1\to\tau^+\mu^-).$$

Notice that the expression (31) does not practically depend on the
lepton masses. Therefore, all discrepancy between the branching
ratios of the decays $Z_1\to\tau\mu, Z_1\to\tau e$ and $Z_1\to
e\mu$ is determined exclusively by the values of the mixing angles
in the heavy neutrinos sector.

\section{Conclusion}
In the framework of the LRM the decay $Z\to\tau\mu$ has been
investigated in two flavor approximation. This decay is prohibited
in the SM by virtue of the fact that it goes with the charged
lepton flavor violation (CLFV). The obtained branching ratio of
this decay does not equal to zero only at the existence of the
neutrino mixings and at the absence of masses degeneracy in the
heavy neutrino sector. From it follows that within the LRM
nonconservation both of neutral and of charged lepton flavors has
the same nature, namely, it is caused by the neutrinos mixing. As
a result, elucidation of the decays $Z\to l_i\overline{l}_k$
($i\neq k$) could provide data concerned the neutrino sector
structure of the LRM. The neutrino sector parameters which could
be measured in that case are as follows: (i) difference of the
heavy neutrino masses; (ii) heavy-heavy neutrino mixing; (iii)
heavy-light neutrino mixing. Note, that information about these
parameters may be also obtained under investigation of the CLFV
Higgs boson decays \cite{Bom18}.

Using the maximal value of the heavy-light neutrino mixing angle
$\varphi$, which was found in collider experiments \cite{CC13}, we
have get the upper bound on the branching ratio of the decay
$Z\to\tau\mu$. The obtained expression appears to be less on two
order of magnitude than the upper bound $\mbox{BR}(Z\to
\tau\mu)_{exp}<1.2\times10^{-5}$ arrived by the experiments at
ATLAS and CMS. However, it is well to bear in mind that this
quantity is not the measured value of the branching ratio. It is
nothing but the precision limit of the current experiments. In
actual truth, the observed value of $\mbox{BR}(Z\to \tau\mu)$ may
prove to be less than $10^{-5}$. As a consequence the experiments
on looking for the CLFV $Z$ boson decays with higher precision
than at present will certainly be continued during the new LHC
runs and at future leptonic colliders where the more high
statistics of $Z$ boson events will be achieved. For example, the
future LHC runs with $\sqrt{s}=14$ TeV and total integrated
luminosity of first 300 $\mbox{fb}^{-1}$ and later 3000
$\mbox{fb}^{-1}$ expect the production of about $10^{10}$ and
$10^{11}$ of the $Z$ boson events, respectively. These large
numbers provide an upgrading of sensitivities to $\mbox{BR}(Z\to
l_k\overline{l}_m)$ of at least one order of magnitude with
respect to the present sensitivity. However, the best
sensitivities for these CLFV decays are expected from next
generation of lepton colliders such as the International linear
collider \cite{HBaer}, Future Circular $e^+e^-$ Collider (FCC-$ee$
--- TLEP) \cite{MB14}, Circular Electron-Positron Collider (CEPC)
\cite{CEPS}, in so far as they can work as $Z$ factory with a very
clean environment. For example, at TLEP \cite{AB14}, where up to
$10^{13}$ $Z$ bosons would be produced, the sensitivities to CLFV
$Z$ decay rates could be improved up to $10^{-13}$.


\begin{thebibliography}{xxxx}
\bibitem{AMB16}A.M.Baldini {\it{et al.}} (MEG), Eur. Phys. J. C {\bf{76}},
434 (2016).
\bibitem{WHB}W.H.Bertl {\it{et al.}} (SINDRUM II), Eur. Phys. J. C
{\bf{47}}, 337 (2006) 176.
\bibitem{AA13}A. Alekou {\it{et al.}}, in Proceedings, 2013 Community
Summer Study on the Future of U.S. Particle Physics: Snowmass on
the Mississippi (CSS2013): Minneapolis, MN, USA, July 29-August 6,
2013 (2013).
\bibitem{LEP1}R. Akers {\it{et al.}} (OPAL), Z. Phys. C {\bf{67}}, 555
(1995).
\bibitem{LEP2}P. Abreu {\it{et al.}} (DELPHI), Z. Phys. C {\bf{73}}, 243
(1997).
\bibitem{ATLAS3}G. Aad {\it{et al.}} (ATLAS), Eur. Phys. J. C
{\bf{77}}, 70 (2017), arXiv:1604.07730 [hep-ex].
\bibitem{GA14}G. Aad {\it{et al.}} (ATLAS),
CERN-EP-2018-052, arXiv:1804.09568 [hep-ex].
\bibitem{VK15}V. Khachatryan {\it{et al.}} [CMS Collaboration],
Phys. Lett. B {\bf{749}}, 337 (2015), arXiv:1502.07400 [hep-ex].
\bibitem{CMS17}CMS Collaboration, Search for lepton flavour
violating decays of the Higgs boson to $\mu\tau$ and $e\tau$ in
proton-proton collisions at $\sqrt{s}=13$ TeV, (2017),
CMS-PAS-HIG-17-001.
\bibitem{KC16} K.Cheung, W.Y.Keung and P.Y.Tseng, Phys. Rev. D
{\bf{93}}, 015010 (2016).
\bibitem{SB16} S.Baek and K.Nishiwaki, Phys. Rev. D {\bf{93}}, 015002
(2016).
\bibitem{Eah16} E. Arganda {\it{et al.}}, Phys. Rev. D {\bf{93}}, 055010
(2016).
\bibitem{AA16} A. Abada {\it{et al.}}, JHEP {\bf{1602}}, 083
(2016).
\bibitem{AAV15}A. Abada {\it{et al.}}, JHEP {\bf{04}}, 051 (2015),
arXiv:1412.6322 [hep-ph].
\bibitem{MAP4}M. A. Perez {\it{et al.}},Int. J. Mod. Phys. A
{\bf{19}}, 159 (2004), arXiv:hep-ph/0305227 [hep-ph].
\bibitem{SDS12}S. Davidson, S. Lacroix, and P. Verdier, JHEP
{\bf{09}}, 092 (2012), arXiv:1207.4894 [hep-ph].
\bibitem{ICP74} J.C.Pati and A.Salam, Phys. Rev. D {\bf{10}}, 275
(1974); R.N.Mohapatra and J.C.Pati, Phys. Rev. D {\bf{11}}, 566
(1975); G.Senjanovic and R.N.Mohapatra, Phys. Rev. D {\bf{12}},
1502 (1975).
\bibitem{Bom97}G.G.Boyarkina, O.M.Boyarkin, Physics of Atomic
Nuclei, {\bf{60}},  601 (1997).
\bibitem{Bom04}O.M.Boyarkin, G.G.Boyarkina, and T.I.Bakanova,
Phys. Rev. D {\bf{70}}, 113010-1 (2004).
\bibitem{Bom18}O.M.Boyarkin, G.G.Boyarkina,D.S.Vasileuskaya,
Int. J. Mod. Phys. A {\bf{33}}, 1850103 (2018).
\bibitem{OMB11}O.M.Boyarkin, {\it{Advanced Particles Physics,
Volume II}} (Taylor and Francis Group, New York, 2011), 555 pp.
\bibitem{ATC17}ATLAS Collaboration Phys.Rev. D {\bf{96}}, 052004
(2017), [arXiv:1703.09127].
\bibitem{TB18}T.Bandyopadhyay {\it{et al.}}, arXiv:1803.07989
[hep-ph].
\bibitem{PSB13}P.S.Bhupal Dev, Chang-Hun Lee, R.N.Mohapatra, Phys. Rev. D
{\bf{88}}, 093010 (2013).
\bibitem{SD16}S.Dell'Oro {\it{et al.}}, Adv. High Energy Phys.
{\bf{2016}}, 2162659 (2016), arXiv:1601.07512 [hep-ph].
\bibitem{CC13}Chien-Yi Chen, P.S.Bhupal Dev, and R.N.Mohapatra,
Phys.Rev. D {\bf{88}}, 033014 (2013).

\bibitem{HBaer}H. Baer {\it{et al.}}, arXiv:1306.6352 [hep-ph].
\bibitem{MB14}M. Bicer {\it{et al.}} [TLEP Design Study Working
Group], JHEP {\bf{1401}}, 164 (2014) doi:10.1007/JHEP01(2014)164
[arXiv:1308.6176 [hep-ex]].
\bibitem{CEPS}CEPC-SPPC Study Group, IHEP-CEPC-DR-2015-01,
IHEP-TH-2015-01, IHEP-EP-2015-01.
\bibitem{AB14}A. Blondel, E. Graverini, N. Serra, and M. Shaposhnikov
(FCC-ee study Team), in Proceedings, 37th International Conference
on High Energy Physics (ICHEP 2014): Valencia, Spain, July 2-9,
2014 , Vol. 273-275 (2016) pp. 1883, arXiv:1411.5230 [hep-ex].
\end{thebibliography}
\end{document}